\documentclass[letterpaper,twocolumn,aps,prl,superscriptaddress,amsmath,amssymb,floatfix]{revtex4-2}
\usepackage{newtxtext,newtxmath}
\usepackage[latin9]{inputenc}

\usepackage{color}
\usepackage{amsmath}
\usepackage{amssymb}
\usepackage{graphicx}
\usepackage{esint}
\usepackage{comment}
\usepackage[unicode=true,
 bookmarks=true,bookmarksnumbered=false,bookmarksopen=false,
 breaklinks=false,pdfborder={0 0 1},backref=false,colorlinks=true]
 {hyperref}
\hypersetup{
 linkcolor=magenta,urlcolor=blue,citecolor=blue,pdfstartview={FitH},hyperfootnotes=false}

\makeatletter

\usepackage{textcomp}
\usepackage{epstopdf}

\pdfpageheight\paperheight
\pdfpagewidth\paperwidth

\@ifundefined{textcolor}{}{%
 \definecolor{BLACK}{gray}{0}
 \definecolor{WHITE}{gray}{1}
 \definecolor{RED}{rgb}{1,0,0}
 \definecolor{GREEN}{rgb}{0,1,0}
 \definecolor{BLUE}{rgb}{0,0,1}
 \definecolor{CYAN}{cmyk}{1,0,0,0}
 \definecolor{MAGENTA}{cmyk}{0,1,0,0}
 \definecolor{YELLOW}{cmyk}{0,0,1,0}
}

\usepackage{xcolor}
\usepackage{soul}
\newcommand{\ket}[1]{\ensuremath{\left|#1\right\rangle}}

\definecolor{blue}{rgb}{0,0,1}
\definecolor{red}{rgb}{1,0,0}
\definecolor{green}{rgb}{0,1,0}

\usepackage{soul}

\makeatother
\begin{document}

\title{Experimental realization of quantum-computing-enhanced sensing}

\author{Yifang Xu}
\thanks{These authors contributed equally to this work.}

\author{Hongwei Huang}
\thanks{These authors contributed equally to this work.}

\author{Yilong Zhou}
\thanks{These authors contributed equally to this work.}

\author{Yunlai~Zhu}
\author{Chuanlong~Ma}
\author{Lida~Sun}
\author{Lintao~Xiao}
\author{Ziyue~Hua}
\author{Weiting~Wang}

\affiliation{Center for Quantum Information, Institute for Interdisciplinary Information Sciences, Tsinghua University, Beijing 100084, China}

\author{Zijie~Chen}
\author{Weizhou~Cai}
\affiliation{Laboratory of Quantum Information, University of Science and Technology of China, Hefei 230026, China}
\affiliation{Hefei National Laboratory, Hefei 230088, China}

\author{Hekang~Li}
\affiliation{School of Physics and ZJU-Hangzhou Global Scientific and Technological Innovation Center, Zhejiang University, Hangzhou 310027, China}
\affiliation{Hefei National Laboratory, Hefei 230088, China}
\author{Haohua~Wang}
\affiliation{School of Physics and ZJU-Hangzhou Global Scientific and Technological Innovation Center, Zhejiang University, Hangzhou 310027, China}
\affiliation{Hefei National Laboratory, Hefei 230088, China}

\author{Xiaoting~Wang}
\affiliation{Institute of Fundamental and Frontier Sciences, University of Electronic Science and Technology of China, Chengdu, 610051, China}

\author{Ming~Li}
\email{lmwin@ustc.edu.cn}
\affiliation{Laboratory of Quantum Information, University of Science and Technology of China, Hefei 230026, China}
\affiliation{Hefei National Laboratory, Hefei 230088, China}

\author{Chang-Ling~Zou}
\email{clzou321@ustc.edu.cn}
\affiliation{Laboratory of Quantum Information, University of Science and Technology of China, Hefei 230026, China}
\affiliation{Hefei National Laboratory, Hefei 230088, China}

\author{Luyan~Sun}
\email{luyansun@tsinghua.edu.cn}
\affiliation{Center for Quantum Information, Institute for Interdisciplinary Information Sciences, Tsinghua University, Beijing 100084, China}
\affiliation{Hefei National Laboratory, Hefei 230088, China}

\begin{abstract}
Quantum algorithms offer computational advantages, yet incorporating them into quantum sensing and converting these advantages into enhanced information acquisition remains challenging. 
Here, we realize quantum-computing-enhanced sensing in a spin-oscillator architecture, where a microwave cavity provides a high-dimensional quantum register and a coupled superconducting qubit serves as the sensor. The unknown signal directly generates the phase oracle operation, establishing a natural physical interface between quantum computation and sensing. We experimentally demonstrate the first Grover search in a bosonic mode and observe quantum amplitude amplification in a Hilbert space spanning more than $128$ photons. 
For the same number of sensing iterations, our protocol extracts more information and resolves more frequency candidates than classical sequential search.
Our results establish signal-driven oracles as a route for harnessing oracle-based quantum algorithms and realizing quantum-computing-enhanced sensing.
\end{abstract}

\maketitle

\noindent
Quantum computation~\cite{Nielsen2010Quantum} and quantum sensing~\cite{Giovannetti2004ScienceQuantum,Degen2017Rev.Mod.Phys.Quantum} are two major applications of quantum technology, each offering distinctive advantages over their classical counterparts~\cite{Huang2025arXiv2508.05720vast}. Beyond leveraging quantum resources to improve the sensing precision in conventional quantum sensing schemes, emerging studies have begun to incorporate quantum algorithms into sensing protocols such as quantum phase estimation~\cite{Khan2025arXiv2507.16918Quantum,Sen2026arXiv2608.17400Quantum}. This naturally raises the intriguing question of whether the computational advantage of quantum algorithms can be directly converted to an advantage in quantum sensing. 
A recent proposal~\cite{Allen2025arXiv2501.07625Quantum} suggests that a quantum processor implementing Grover's search algorithm~\cite{Grover1997Phys.Rev.Lett.Quantum} can accelerate the identification of unknown signal frequency, providing a quadratic speedup in the number of interrogations required for frequency search. 
When combined with quantum-enhanced amplitude estimation~\cite{Huang2024Appl.Phys.Rev.Entanglement,Wang2019Nat.Commun.Heisenberg,Pan2025PRXQuantumRealization}, this forms the Grover-Heisenberg sensing framework, as relevant to searches for gravitational waves~\cite{Tse2019Phys.Rev.Lett.Quantum} and oscillating dark-matter fields~\cite{Backes2021Naturequantum,Zheng2026PRLQuantum}.
However, an experimental realization of such quantum-computing-enhanced sensing remains to be demonstrated.

The incorporation of quantum algorithms into quantum sensing protocols introduces a fundamental challenge not present in conventional demonstrations of quantum algorithms. In previous proof-of-principle implementations of quantum algorithms, such as Grover search, the target-dependent oracle is typically realized via specially designed quantum control operations programmed for a predetermined target state~\cite{Figgatt2017Nat.Commun.Complete,Pokharel2024npjQuantumInf.Better,AbuGhanem2025Sci.Rep.Characterizing,Main2025NatureDistributed}. In quantum sensing, however, the physical signal to be identified is unknown, and thus the corresponding oracle cannot be programmed in advance. Therefore, the core requirement is to realize a signal-driven oracle, in which the unknown signal itself participates in or even directly drives the execution of the quantum algorithm. In this framework, each coherent interaction with the signal effectively constitutes an oracle, enabling quantum amplitude amplification to steer the system toward distinguishable quantum states.

Bosonic quantum systems provide a natural architecture to realize such a signal-driven interface between quantum computating and quantum sensing. Their high-dimensional Fock space offers a scalable computational Hilbert space~\cite{Ofek2016NatureExtending,Hu2019Nat.Phys.Quantum,CampagneIbarcq2020NatureQuantum,Cai2021FundamentalResearchBosonic}, while the coupling to ancillary nonlinear elements support both universal quantum control~\cite{Heeres2015Phys.Rev.Lett.Cavity,Krastanov2015Phys.Rev.AUniversal,Heeres2017Nat.Commun.Implementing,Eickbusch2022Nat.Phys.Fast,Chen2025Sci.Adv.Robust} and implementations of quantum algorithms~\cite{Knill_PRL1998,Wang_PRL2017} and quantum sensing protocols~\cite{Deng2024Nat.Phys.Quantum,Pan2025PRXQuantumRealization,Xu2026Nat.Phys.Principles,Hua2026arXiv2602.23254Quantum,Zheng2026PRLQuantum}.
In a cavity-qubit system, different Fock states induce distinct resonance frequencies of the coupled qubit~\cite{Schuster2007NatureResolving}, thereby establishing a one-to-one correspondence between the Fock-state search space and candidate sensing frequencies. 

Here, we exploit this correspondence to experimentally demonstrate Grover-enhanced frequency sensing in a superconducting cavity-qubit system~\cite{Wallraff2004NatureStrong,Reagor2016Phys.Rev.BQuantum,Blais2021Rev.Mod.Phys.Circuit,Milul2023PRXQuantumSuperconducting}. 
We realize the first experimental implementation of Grover search in a bosonic mode and demonstrate deterministic preparation of Fock states through quantum amplitude amplification, achieving Grover amplification up to 128 photons. By allowing the unknown frequency signal to directly generate the Grover oracle, the quantum processor extracts more information and resolves a larger number of frequency candidates than classical sequential search, exhibiting the characteristic quadratic scaling of Grover search. Our work provides the first experimental evidence that quantum computation can serve as an active resource for quantum sensing, establishing a new route toward integrating quantum computation with quantum sensing.

\begin{figure*}[tbp]
    \centering
    \includegraphics{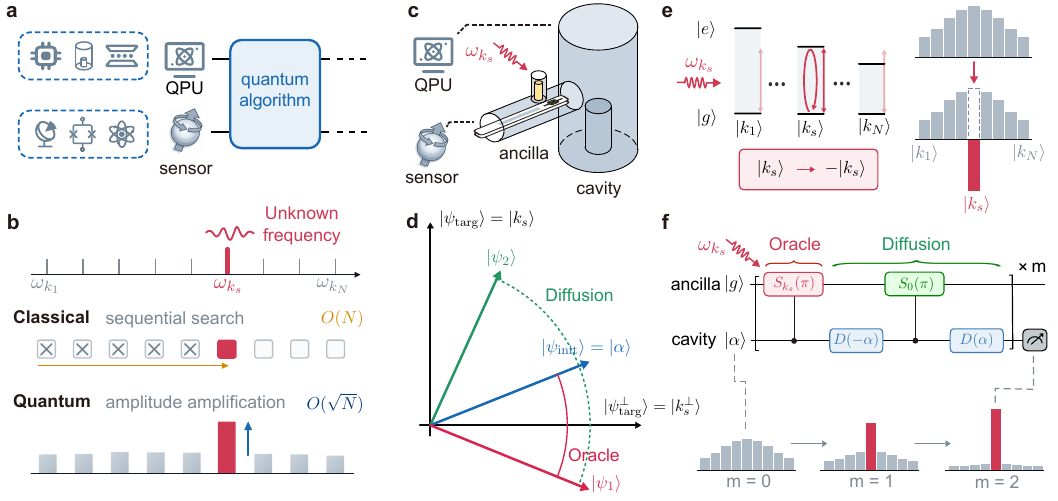}
    \caption{\textbf{Concept of quantum-computing-enhanced sensing in a bosonic system.}
    \textbf{a}, Concept. Running quantum algorithms in a coupled QPU-sensor system offers the potential to accelerate sensing processes.
    \textbf{b}, Sensing task and strategies. The goal is to identify an unknown frequency $\omega_{k_s}$ from a discrete set $\{\omega_k\}$ with $N$ candidates. Classical sequential search has complexity $O(N)$, while quantum Grover search has complexity $O(\sqrt{N})$.
    \textbf{c}, Experiment device. In our setup, the bosonic cavity provides the high-dimensional computational space of the QPU, while the coupled qubit serves as the quantum sensor for receiving the target signal and also acts as an ancilla for nonlinear cavity control.
    \textbf{d}, Grover algorithm schematic. The initial state $\ket{\psi_\mathrm{init}}=\ket{\alpha}$ (blue) is decomposed into the target state $\ket{\psi_\mathrm{targ}}=\ket{k_s}$ and an orthogonal state $|\psi_\mathrm{targ}^{\perp}\rangle=\ket{k_{s}^\perp}$. The oracle flips the phase of the target component, yielding $|\psi_1\rangle$ (red). The diffusion operator then reflects about $\ket{\alpha}$, producing $|\psi_2\rangle$ (green). Repeating this sequence drives the state toward the target $\ket{k_s}$. 
    \textbf{e}, Oracle implementation. The oracle exploits the number-resolved cavity-qubit interaction. Since the qubit transition frequency depends on the cavity Fock number, the unknown signal frequency $\omega_{k_s}$ selectively addresses the qubit only when the cavity occupies $\ket{k_s}$. This conditional qubit rotation imparts a $\pi$ geometric phase on $\ket{k_s}$, thereby realizing the oracle. 
    \textbf{f}, Pulse sequence. The complete pulse sequence for the quantum-computing-enhanced sensing protocol is shown.
    }
\label{fig:concept}
\end{figure*}

\bigskip
\noindent\textbf{Signal-driven Grover sensing in a bosonic system}

\noindent
The concept of quantum-computing-enhanced sensing is illustrated in Fig.~\ref{fig:concept}a. The system consists of a quantum processing unit (QPU) and a sensor, where the sensor directly interacts with the unknown signal and mediates a signal-dependent operation on the QPU.
We consider the sensing task of identifying an unknown signal frequency from a discrete set of candidates, $\{\omega_k:k=k_1,k_2,\dots,k_N\}$, as illustrated in Fig.~\ref{fig:concept}b (top). 
The classical strategy performs a sequential search, where each candidate frequency is tested individually until the input signal triggers the sensor [Fig.~\ref{fig:concept}b (middle)], with the required number of steps scaling as $O(N)$. 
In contrast, by converting the input signal with unknown frequency into a phase oracle operation that flips the phase of the corresponding quantum state in the QPU, we can turn each interrogation of the signal into a single oracle call and thus compute oracle-based quantum algorithms to implement sensing. Taking Grover's algorithm as an example [Fig.~\ref{fig:concept}b (bottom)], the signal-driven oracle amplifies the target population and enables rapid identification of the unknown signal frequency in $O(\sqrt{N})$ interrogations. Since this work focuses on the enhancement of frequency identification by quantum computation, we assume, without loss of generality, that the signal amplitude and interrogation time are known.
 
We implement the quantum-computing-enhanced sensing in a superconducting circuit quantum electrodynamics platform~\cite{Blais2021Rev.Mod.Phys.Circuit}, which offers the number-resolved sensor response required by the signal-driven oracle. Our experimental system consists of a three dimensional superconducting cavity coupled to a transmon qubit~\cite{Koch2007Phys.Rev.ACharge}, as shown in Fig.~\ref{fig:concept}c. The bosonic cavity serves as a high-dimensional quantum register, while the coupled qubit directly interacts with the signal.
The device is specially designed~\cite{Zhou2026inpreparation} so that the cavity exhibits a strong cross-Kerr interaction with the qubit while maintaining an extremely weak self-Kerr nonlinearity.
Owing to the number-resolved interaction arising from the cross-Kerr coupling, different cavity Fock states $|k\rangle$ correspond to distinct qubit transition frequencies $\omega_k$, establishing a one-to-one mapping between the Fock states and the candidate frequencies. By initializing the cavity in a superposition state $|\psi_{\mathrm{init}}\rangle = \sum_k c_k |k\rangle$ and the sensor qubit in the ground state $|g\rangle$, the sensing problem is thereby converted into identifying the unknown input frequency from the final Fock state component.

\begin{figure*}[t]
    \centering
    \includegraphics{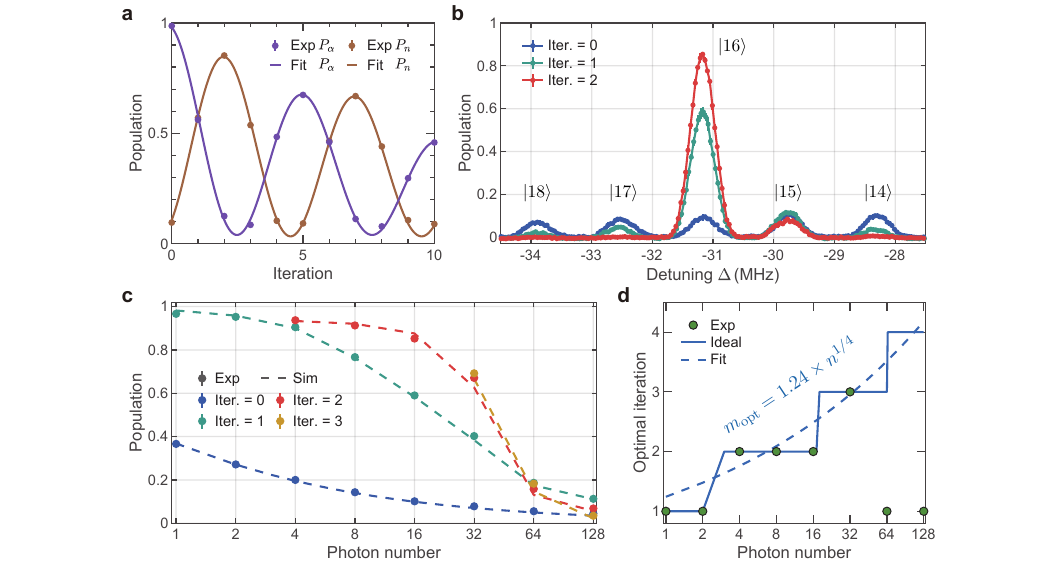}
    \caption{\textbf{Results of Grover search.} 
     \textbf{a}, Population of the cavity state projected onto the $|\alpha=3.86\rangle$ and $|n=16\rangle$ axes, confirming Rabi oscillation in the subspace $S=\{|\alpha\rangle, |n\rangle\}$. 
     \textbf{b}, Number-splitting experiments of the cavity photon number for different Grover iterations at $n=16$. After two iterations, the population of $\ket{16}$ reaches a maximum of 85.4\%. 
     \textbf{c}, Performance of the Grover algorithm in searching for the target Fock state $\ket{n}$. Color dots (dashed lines) represent experimental (simulated) results for different iteration counts. Error bars are smaller than the marker sizes and omitted. 
     \textbf{d}, Optimal number of iterations versus Fock number. Dots represent experimental data, and lines show the theoretical expectation.
    }
\label{fig:rabi}
\end{figure*}

Figure~\ref{fig:concept}d shows a schematic of the Grover algorithm, which requires two basic operations: the oracle and the diffusion operators. We take a coherent state as the initial state, for reasons discussed below. The key ingredient of quantum-computing-enhanced sensing is that the unknown signal itself generates the oracle. Because the qubit resonance depends on the cavity photon number, the evolution of the qubit induced by the signal $\omega_k$ is an operation conditioned on the cavity state [Fig.~\ref{fig:concept}e (left)]. The amplitude and duration of the signal are calibrated such that the joint system undergoes a cyclic evolution and acquires a geometric phase of $\pi$ when the unknown signal is resonant with one of the transition frequencies, i.e., $\omega_k=\omega_{k_s}$. The overall transformation
\begin{equation}
    U_{k_s} |\psi_{\mathrm{init}}\rangle\otimes|g\rangle = (I - 2|k_s\rangle\langle k_s|) |\psi_{\mathrm{init}}\rangle \otimes |g\rangle
\end{equation}
thus returns the qubit to its initial state and encodes the signal information as a $\pi$ phase in the corresponding Fock state [Fig.~\ref{fig:concept}e (right)], which is precisely the Grover phase oracle. This oracle implementation shares the same spirit as the selective number-dependent arbitrary phase (SNAP) gate~\cite{Krastanov2015Phys.Rev.AUniversal,Heeres2015Phys.Rev.Lett.Cavity}, but differs in that the phase oracle is now determined by the unknown signal rather than being pre-programmed for a predetermined target state as in a conventional SNAP gate. This distinction constitutes the essential step that converts Grover search from a computational routine into a sensing protocol.

The complete execution of the Grover algorithm also requires a diffusion operation $U_\mathrm{D} = 2|\psi_\mathrm{init}\rangle\langle\psi_\mathrm{init}| - I$, which flips the phase of the initial state up to a global phase. While the standard Grover search requires an initial state with a uniform distribution, such a state is not native to bosonic systems, nor is the diffusion operation.
We therefore employ a coherent state $\ket{\alpha}$ as the initial state, which can be directly prepared from vacuum by a displacement operation $D(\alpha)$.
More importantly, the corresponding diffusion operation admits a simple realization
\begin{equation}
    U_D=D(\alpha) P_0^\pi D(-\alpha) = I - 2|\alpha\rangle\langle \alpha|
\end{equation}
up to a global $\pi$ phase, where $P_0^{\pi}$ is a phase flip on the vacuum state $\ket{0}$ implemented via the SNAP gate. The choice of a coherent state simultaneously simplifies both state initialization and the most demanding state-dependent reflection.
The oracle and the diffusion operators are applied alternately for $m$ iterations, after which the photon-number distribution of the cavity is measured to infer the input frequency [Fig.~\ref{fig:concept}f].

\bigskip
\noindent\textbf{Bosonic Grover dynamics and Fock-state amplification}

\noindent
Before implementing signal-driven Grover sensing, we first validate the underlying amplitude-amplification dynamics of the bosonic QPU using a programmed Fock-state oracle.
For a fixed target $|n\rangle$, the Grover dynamics in the cavity QPU can be understood as a discrete Rabi process within the subspace spanned by $|\alpha\rangle$ and $|n\rangle$.
We first experimentally verify the deterministic Rabi oscillations within this subspace. 
Taking $\alpha =3.86$ and $n=16$ as an example, we measure the population of $\ket{n}$ and $\ket{\alpha}$ simultaneously, as illustrated in Fig.~\ref{fig:rabi}a. The out-of-phase oscillation between the two curves confirms the presence of Rabi oscillations within the subspace $S=\{|\alpha\rangle, |n\rangle\}$.

Due to the non-uniform photon-number distribution of the coherent state, the required number of Grover iterations is not the same and is determined by $m=\left\lceil\left[(\pi/2)/\sin^{-1}(|\langle \alpha|n\rangle|)-1\right]/2\right\rceil$ for different target Fock states.
There is considerable freedom in adjusting the parameters to maximize the probability amplification of the target Fock state toward unity.
By slightly tuning $\alpha$ away from $\sqrt{n}$, the trajectory reaches the target state precisely after an integer number of algorithm iterations.
As shown in Fig.~\ref{fig:rabi}b, after two iterations of Grover search, the population of $|16\rangle$ reaches a maximum of 85.4\%. In the measured Fock space, aside from $|16\rangle$, only $|15\rangle$ shows a noticeable population, which may arise from transmon decay during the SNAP operation or the measurement after the Grover algorithm.

To demonstrate the scalability of our method, we measure the population of prepared Fock states with different photon numbers, as shown in Fig.~\ref{fig:rabi}c. For different Fock states, our experiment achieves fidelities exceeding 90\% for small photon numbers. Using a fidelity of 50\% as a reference threshold, the preparation can be scaled to over 32 photons. Even when extended to 128 photons, the amplification effect of the Grover algorithm remains observable. 

A clear degradation of the experimental performance appears around 32 to 64 photons, which we attribute to the influence of the self-Kerr nonlinearity. During the diffusion operations associated with the protocol, when preparing a Fock state $|n\rangle$, the photon number temporarily spreads over a Fock space ranging from $0$ to about $4n$. In our device, the effective suppression of self-Kerr is mainly achieved within $128$ photons (see Supplementary Materials~\cite{supplement}). {This Kerr-induced distortion is further reflected in Fig.~\ref{fig:rabi}d, which plots the optimal number of iterations versus the Fock number. While the ideal scaling follows $O(n^{1/4})$~\cite{Jin2026arXiv2602.12156Deterministic,Roy2026arXiv2608.23389Exact}, the experimental data deviate markedly at high photon numbers, where additional iterations exacerbate performance degradation due to accumulated algorithmic imperfections.}

The experimental results indicate that the core hardware requirement for bosonic Grover search is a strong number-selective interaction, enabled by a large cross-Kerr strength, and simultaneously a weak intrinsic cavity nonlinearity for high-fidelity oracle and diffusion operations.
In the future, employing an ancilla qubit with stronger nonlinearity, such as a fluxonium~\cite{Manucharyan2009ScienceFluxonium}, may further suppress higher-order effects that emerge at larger photon numbers.

\begin{figure}[tbp]
    \centering
    \includegraphics{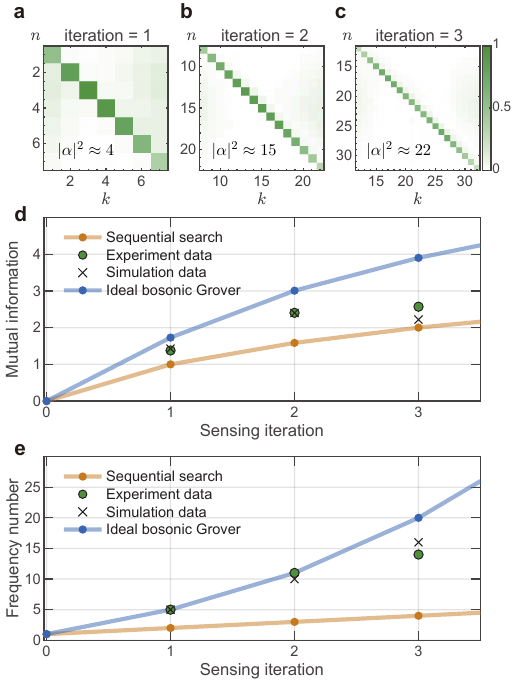}
    \caption{\textbf{Results of Grover-enhanced sensing.} 
    \textbf{a-c}, Experimental cavity photon-number distributions versus input signal frequency, for one-, two-, and three-iteration Grover sequences, respectively. The optimal initial coherent-state sizes are $|\alpha|^2 \approx 4$, $15$, and $22$, respectively. Columns correspond to different probe frequencies, and the vertical axis within each column indicates the population distribution of cavity photon states after the Grover sequence. A strong correlation is observed between the input signal and the final cavity state. \textbf{d}, Extracted MI from experiment data (green dots), compared with simulation results (black cross), the classical sequential-search bound (orange), and the Grover protocol in this work (blue). For a given number of sensing iterations, the Grover-enhanced approach yields higher MI than the sequential scheme. 
    \textbf{e}, Number of resolvable candidate frequencies at maximum MI versus sensing iterations, representing the sensing bandwidth. {The Grover-enhanced approach outperforms the sequential scheme by resolving a larger set of frequencies.}
    }
\label{fig:sensing}
\end{figure}

\bigskip
\noindent\textbf{Grover-enhanced frequency sensing}

\noindent
Having realized the Grover algorithm in our bosonic system, we extend its application to quantum sensing by employing the same operation sequence to estimate the signal frequency.
Candidate signal frequencies are encoded into oracle drives, and we measure the conditional photon-number distribution of the cavity after $m$ Grover sensing iterations for each input frequency $\omega_k$. As shown in Figs.~\ref{fig:sensing}a-c,
a pronounced correlation is clearly observed between the final cavity states and the input signal frequencies. These dataset can be conceptualized as a matrix, where distinct columns represent input signals of varying frequencies $\omega_k$, and the elements within a given column delineate the cavity photon-number distribution following the execution of the Grover sequence. Ideally, the protocol maps each candidate signal frequency onto a distinct Fock-state outcome, rendering a perfectly diagonal identity matrix and enabling the input frequency to be inferred from a single measurement of the cavity. 

In practice, however, deviations from this ideal behavior arise. The primary imperfection stems from the non-uniform probability amplitude distribution of the Fock components within the initial coherent state $|\alpha\rangle$. The Rabi frequency of the Grover algorithm depends on the overlap between the initial state and the target component. Consequently, under a fixed number of Grover iterations, target states with excessively large initial amplitudes undergo over-rotation, whereas those with insufficient initial amplitudes remain under-amplified. Additional deviations are partially attributable to cavity and qubit dissipation as well as cavity nonlinearities.
This behavior elucidates the spatial profile of the diagonal elements observed in the matrix. 
For example, in Fig.~\ref{fig:sensing}c, the diagonal population exhibits a suppressed valley near state $|22\rangle$, subsequently increases as the distance from state $|22\rangle$ increases, and ultimately tapers off at the boundaries.

The sensing advantage of our protocol can be characterized by two quantities. 
The first is the mutual information (MI) between the input frequency and the measured cavity state, obtained under optimized operating conditions for a fixed number of sensing iterations $m$. MI measures the information extracted from the system after the sensing measurement, equivalently quantifying the reduction in uncertainty about the input frequency candidates. The definition of MI and the calculation of its upper bounds for classical and quantum strategies are described in Supplementary Materials~\cite{supplement}.
The second is the number of resolvable candidate frequencies, $N$, under a fixed number of optimized sensing iterations $m$, which defines the effective range of the search space, i.e., the sensing bandwidth.

In our system, maximizing the MI for a given $m$ requires optimizing the amplitude of the initial coherent state. By systematically scanning this parameter, we determine the optimal initial coherent-state sizes to be approximately $|\alpha|^2 = 4$, $15$, and $22$ for one, two, and three Grover iterations, respectively. 
It is worth noting that, for a fixed $m$, achieving the maximum MI also requires calibrating the frequency range $N$. A simple example is provided by classical sensing: a straightforward strategy is to sequentially scan all candidate frequencies, where $m$ measurements suffice to fully resolve a task containing $N=m+1$ candidates. In this case, the obtained MI reaches the classical upper bound, as shown by the orange curve in Fig.~\ref{fig:sensing}d. A similar optimization is required for the quantum strategy. Specifically, we use a simulated annealing algorithm to identify a subset with the largest MI within a sufficiently large candidate-frequency space (see Supplementary Materials~\cite{supplement}).

The corresponding MI values and resolvable frequency candidate numbers $N$ for $m=1,\,2,\,3$ are shown as green points in Figs.~\ref{fig:sensing}d and \ref{fig:sensing}e, respectively. The blue curves represent the dissipation-free limit of our protocol with coherent-state inputs, providing the theoretical upper bounds for the present Grover-algorithm scheme. 
The crosses are the numerical results considering system dissipation based on experimentally calibrated device parameters. The remaining deviations may arise from differences between the inferred and actual Hamiltonians and from applying vacuum-calibrated Bayesian correction parameters at large photon numbers.
In addition, the experimentally optimized SNAP-gate durations are not necessarily optimal for the inferred Hamiltonian used in the simulations, which may degrade simulation performance at large photon numbers (corresponding to $m=3$ iterations).Further details on the data processing and numerical simulations can be found in the Supplementary Materials~\cite{supplement}.

Compared with the classical sequential-search strategy (orange curves in Figs.~\ref{fig:sensing}d and \ref{fig:sensing}e), our protocol outperforms on two fronts. For the same number of signal interrogations, $m=2$, it resolves a search space nearly four-times-larger (11 vs 3 candidates) and enhances the extracted information from 1.58 to 2.40 bits. Our approach thus surpasses the classical strategy in both sensing bandwidth and information gain, clearly demonstrating the advantage of quantum-computing-enhanced sensing.

\bigskip
\noindent\textbf{Discussion}

\noindent
In summary, we experimentally realize quantum-computing-enhanced sensing by designing the oracle of a quantum algorithm to be generated by the unknown signal. Rather than processing measurement outcomes after the sensing task is completed, the operation of the quantum processor is directly driven by the signal itself, which coherently amplifies the population of the corresponding state outcome. The significance of this approach is not restricted to Grover search. The signal-driven oracle establishes a general physical interface between quantum sensing and a broader class of oracle-based quantum algorithms~\cite{Deutsch1992ProceedingsMathematicalandPhysicalSciencesRapid,Montanaro2016Phys.Rev.AQuantum}. When an unknown physical parameter can coherently generate the oracle required by a quantum algorithm, the corresponding quantum speedup may in principle be converted into an enhancement in sensing or information acquisition.
This perspective extends the role of quantum computation from executing a particular sensing sequence to enhancing information acquisition by quantum algorithms.

The present experiment isolates the frequency component of a signal from the amplitude, in which the signal amplitude is calibrated to realize the desired phase oracle. It should be emphasized that many conventional quantum sensing protocols have been developed to enhance the precision for estimating an unknown signal amplitude~\cite{Giovannetti2004ScienceQuantum,Wang2019Nat.Commun.Heisenberg,Marciniak2022NatureOptimal,Li2023ScienceImproving,Mao2023Nat.Phys.Quantum,Huang2024Appl.Phys.Rev.Entanglement,Deng2024Nat.Phys.Quantum,Pan2025PRXQuantumRealization,Cai2025Nat.Commun.Quantum,Hua2026arXiv2602.23254Quantum}.
A natural extension is to combining the computing-enhanced Grover speedup and Heisenberg-limited quantum amplitude sensing to simultaneously measure the unknown frequency and amplitude of a signal toward the broader Grover-Heisenberg limit~\cite{Allen2025arXiv2501.07625Quantum}. This could potentially be realized by introducing multiple qubits or a bosonic mode as the sensor. 

Our results also reveal several distinctive features of bosonic quantum processors for implementing oracle-based quantum algorithms. Previous experimental demonstrations of Grover search have primarily relied on multi-qubit architectures~\cite{Figgatt2017Nat.Commun.Complete,Pokharel2024npjQuantumInf.Better,AbuGhanem2025Sci.Rep.Characterizing,Main2025NatureDistributed}, where the oracle operation becomes increasingly difficult to realize as the system size grows. In contrast, the spin-oscillator architecture is hardware efficient and naturally realizes the oracle through the Fock-state-dependent frequency response of the ancilla qubits.
Beyond sensing, the demonstrated amplitude-amplification dynamics also provide an algorithmic strategy for preparing and controlling high-photon-number bosonic states~\cite{Jin2026arXiv2602.12156Deterministic,Nagib2025Phys.Rev.Lett.Efficient,Austin2026arXiv2607.14239Fock,Roy2026arXiv2608.23389Exact}.

\bigskip
\noindent\textbf{\large{}{}Acknowledgments}
{\large\par}
\noindent
We thank the USTC Center for Micro and Nanoscale Research and Fabrication. We also acknowledge the Supercomputing Center of USTC.

\smallskip{}
\noindent\textbf{\large{}{}Funding Statement}
{\large\par}
\noindent This work was funded by the National Natural Science Foundation of China [Grants No. 92265210, 12204052, 92365301, 12474498, 12574539, 12550006, 92565301, 12504580, 92265208], Innovation Program for Quantum Science and Technology [Grant No.~2021ZD0300200 and 2024ZD0301500], and the Sichuan Science and Technology Program (Grant No.~2025YFHZ0336). This work was also supported by the Fundamental Research Funds for the Central Universities and USTC Research Funds of the Double First-Class Initiative.


%

\end{document}